\documentstyle[aas2pp4]{article}
\makeatletter
\let\@internalcite\cite
\def\cite{\def\@citeseppen{-1000}%
    \def\@cite##1##2{(##1\if@tempswa , ##2\fi)}%
    \def\citeauthoryear##1##2##3{##1 ##3}\@internalcite}

\def\citeNP{\def\@citeseppen{-1000}%
    \def\@cite##1##2{##1\if@tempswa , ##2\fi}%
    \def\citeauthoryear##1##2##3{##1 ##3}\@internalcite}
\def\citeN{\def\@citeseppen{-1000}%
    \def\@cite##1##2{##1\if@tempswa , ##2)\else{)}\fi}%
    \def\citeauthoryear##1##2##3{##1 (##3}\@citedata}
\def\citeA{\def\@citeseppen{-1000}%
    \def\@cite##1##2{(##1\if@tempswa , ##2\fi)}%
    \def\citeauthoryear##1##2##3{##1}\@internalcite}
\def\citeANP{\def\@citeseppen{-1000}%
    \def\@cite##1##2{##1\if@tempswa , ##2\fi}%
    \def\citeauthoryear##1##2##3{##1}\@internalcite}
\def\shortcite{\def\@citeseppen{-1000}%
    \def\@cite##1##2{(##1\if@tempswa , ##2\fi)}%
    \def\citeauthoryear##1##2##3{##2 ##3}\@internalcite}
\def\shortciteNP{\def\@citeseppen{-1000}%
    \def\@cite##1##2{##1\if@tempswa , ##2\fi}%
    \def\citeauthoryear##1##2##3{##2 ##3}\@internalcite}
\def\shortciteN{\def\@citeseppen{-1000}%
    \def\@cite##1##2{##1\if@tempswa , ##2)\else{)}\fi}%
    \def\citeauthoryear##1##2##3{##2 (##3}\@citedata}
\def\shortciteA{\def\@citeseppen{-1000}%
    \def\@cite##1##2{(##1\if@tempswa , ##2\fi)}%
    \def\citeauthoryear##1##2##3{##2}\@internalcite}
\def\shortciteANP{\def\@citeseppen{-1000}%
    \def\@cite##1##2{##1\if@tempswa , ##2\fi}%
    \def\citeauthoryear##1##2##3{##2}\@internalcite}
\def\citeyear{\def\@citeseppen{-1000}%
    \def\@cite##1##2{(##1\if@tempswa , ##2\fi)}%
    \def\citeauthoryear##1##2##3{##3}\@citedata}
\def\citeyearNP{\def\@citeseppen{-1000}%
    \def\@cite##1##2{##1\if@tempswa , ##2\fi}%
    \def\citeauthoryear##1##2##3{##3}\@citedata}

\def\@citedata{%
	\@ifnextchar [{\@tempswatrue\@citedatax}%
				  {\@tempswafalse\@citedatax[]}%
}

\def\@citedatax[#1]#2{%
\if@filesw\immediate\write\@auxout{\string\citation{#2}}\fi%
  \def\@citea{}\@cite{\@for\@citeb:=#2\do%
    {\@citea\def\@citea{, }\@ifundefined
       {b@\@citeb}{{\bf ?}%
       \@warning{Citation `\@citeb' on page \thepage \space undefined}}%
{\csname b@\@citeb\endcsname}}}{#1}}%

\def\@citex[#1]#2{%
\if@filesw\immediate\write\@auxout{\string\citation{#2}}\fi%
  \def\@citea{}\@cite{\@for\@citeb:=#2\do%
    {\@citea\def\@citea{; }\@ifundefined
       {b@\@citeb}{{\bf ?}%
       \@warning{Citation `\@citeb' on page \thepage \space undefined}}%
{\csname b@\@citeb\endcsname}}}{#1}}%

\def\@biblabel#1{}

\newlength{\bibhang}
\makeatother

\def\ebv{$E($\bv)}
\def\etal{et~al.}
\def\eg{e.g.}
\newcommand{\fig}[1]{Fig.~\ref{#1}}

\def\kms{km s$^{-1}$}

\def\hone{\ion{H}{1}}

\def\iue{{\it IUE}}

\def\orf{ORFEUS}
\def\rv{$R_V$}

\begin{document}

\twocolumn
 
\title{ORFEUS-II Observations of the Ultraviolet-Bright Star\\
Barnard 29 in M13\footnotemark}
\footnotetext{Based on the development and utilization of ORFEUS
(Orbiting and Retrievable Far and Extreme Ultraviolet
Spectrometers), a collaboration of the Institute for Astronomy and
Astrophysics at the University of T\"{u}bingen, the Space
Astrophysics Group of the University of California at Berkeley, and
the Landessternwarte Heidelberg.}

\author{W. Van Dyke Dixon and Mark Hurwitz}
\affil{Space Sciences Laboratory\\
University of California, Berkeley, California 94720-5030\\
vand@ssl.berkeley.edu}

\begin{center}{To appear in {\it The Astrophysical Journal (Letters)}}\end{center}

\begin{abstract}

The UV-bright star Barnard 29 in the globular cluster M13 was observed
for 5300 seconds with the Berkeley spectrometer on the ORFEUS-SPAS II
mission in 1996 November--December. The resulting spectrum extends from
the interstellar cutoff at 912 \AA\ to $\sim$ 1200 \AA\ at a resolution
of $\sim$ 0.33 \AA. It shows numerous absorption features, both
photospheric and interstellar, but no significant emission other than
diffuse emission of local origin.  The Kurucz synthetic stellar
spectrum that best fits the data has $T_{eff} = 21,000$~K, $\log g =
3.0$, and [M/H] = $-2.5$. This effective temperature and surface
gravity are consistent with previous results, but the derived
metallicity is lower than that of other M13 giants, for which [Fe/H] =
$-1.60$.  Using high-resolution synthetic spectra, we determine the
photospheric abundances of C, S, and Fe, species unobservable in the
optical.  We find $\log \epsilon$(C) = $6.15 \pm 0.10$, $\log
\epsilon$(S) = $5.34 \pm 0.50$, and $\log \epsilon$(Fe) =
$5.30^{+0.22}_{-0.26}$.  Again, the Fe abundance is lower than
expected.  This anomaly may reflect selective condensation of metals
onto dust grains at the end of the AGB phase, as has been suggested for
some cooler post-AGB stars with peculiar Fe abundances.\\

\end{abstract}
 

\section{INTRODUCTION}

\renewcommand{\arraystretch}{1.1}

The ultraviolet-bright stars in globular clusters lie in the upper left
corner of the color-magnitude diagram. Brighter than the horizontal
branch (HB) and bluer than the asymptotic giant branch (AGB), this
class of objects includes evolved HB, post-HB, and post-AGB stars.
Fewer than 50 globular cluster UV-bright stars are known; of these, only
13--16 may be classified as post-AGB stars \cite{de_Boer87}. The brief
post-AGB phase (typically $10^4$ years; \citeANP{S83} \citeyearNP{S81,S83}) is one of the least well understood phases of stellar evolution,
yet the nucleosynthesis, dredge-up, and mass-loss processes that occur
during this phase are principally responsible for the enrichment of the
interstellar medium (ISM), and for carbon and oxygen in particular
\cite{WST89}. To improve our understanding of the final stages of 
a star's lifetime, as well as the mechanisms by which it enriches the
ISM, we observed the UV-bright star Barnard 29
with the Berkeley spectrograph on the ORFEUS-SPAS II mission.

Barnard 29 is a well-studied UV-bright star in the globular cluster M13
\cite{SS70,de_Boer85,CDK94}. Stellar and cluster parameters
are presented in Table \ref{params}.  Recently, \shortciteN{CDK94}
used an LTE analysis of high-resolution optical spectra of Barnard 29
to derive its atmospheric parameters and abundances.  The authors found
$T_{eff} = 20,000$ K and $\log g = 3.0$, confirming Barnard 29 as a
post-AGB star.  The star's helium abundance is nearly solar.  Metal
abundances are about 1 dex below solar, except for carbon, which is
deficient by some 2.4 dex, and nitrogen, which shows a much smaller
underabundance of 0.7 dex (relative to solar).  The authors were able
to set only upper limits on the abundance of three important elements,
C, S, and Fe.  We use the far-UV spectrum of Barnard 29 to set
independent constraints on the star's atmospheric parameters and to
determine these elemental abundances.

\nocite{GA86,CM79,Djorgovski93,Stark92,Arp65p401,CDK94,KSSSLP97}

\section{OBSERVATIONS AND DATA REDUCTION}

The Berkeley spectrograph, located at the prime focus of the 1-m ORFEUS
telescope, flew aboard the space shuttle {\it Columbia} on the
ORFEUS-SPAS II mission in 1996 November--December. The spectrograph's
far-UV sensitivity extends from the interstellar cutoff at 912~\AA\ to
about 1220~\AA, with a mean spectral resolution of 95 km s$^{-1}$ FWHM
(about 0.33~\AA) for point sources.  Its effective area peaks at about
9 cm$^2$ near 1000~\AA.  The general design of the Berkeley
spectrograph is discussed by \citeANP{HB86}
\citeyear{HB86,HB96p601}, while its
calibration and performance on the ORFEUS-SPAS II mission are described
by \citeN{ORFEUS98}.

\begin{table}[bt]
\caption{Stellar Parameters\label{params}}
\begin{tabular}{lcc}
\\ [-4mm] \hline\hline
Parameter & Value & Reference \\ \hline
Spectral Type   &       B2p     &       1 \\
$V$             &       13.14   &       2 \\
$B-V$           &       $-0.16$ &       2 \\
$E(B-V)$        &       0.02    &       3 \\
Distance (kpc)  &       7.2     &       3 \\
$\log N_{\rm H}$ (cm$^{-2}$) & 20.17 & 4 \\
$V_{r, {\rm cluster}}$ (\kms) &       $-241 \pm 10$ & 5 \\
$V_{r, {\rm star}}$ (\kms)    &       $-251 \pm 7$  & 6 \\
${\rm [Fe/H]_{cluster}}$  &       $-1.60$ & 7 \\
$T_{eff}$ (K)   &       $20,000 \pm 1000$ & 6 \\
$\log g$ (dex)  &       $3.0 \pm 0.1$ & 6 \\ \hline
\multicolumn{3}{p{225pt}}{{\sc References:}
(1) Garrison \& Albert 1986;
(2) Cudworth \& Monet 1979;
(3) Djorgovski 1993;
(4) Stark \etal\ 1992;
(5) Arp 1965;
(6) Conlon \etal\ 1994;
(7) Kraft \etal\ 1997.}
\end{tabular}
\end{table}

\begin{figure}[tb]
\plotone{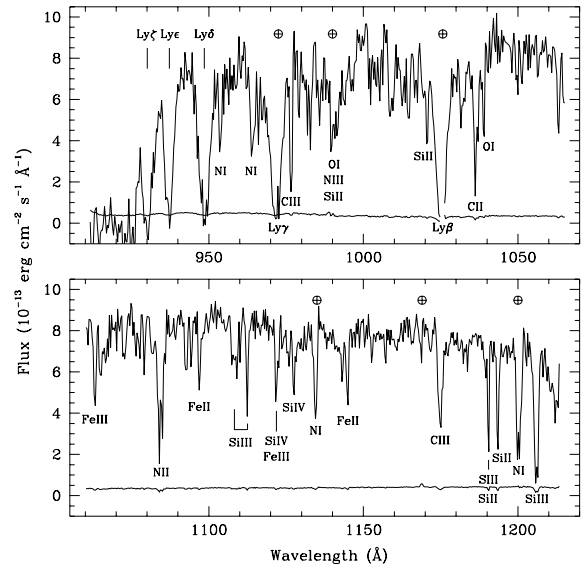}
\caption{Far-UV spectrum of the UV-bright star Barnard 29 in M13
obtained with the Berkeley spectrometer on the \orf\ telescope.  The
data are flux-calibrated, airglow-subtracted, and binned (for this
figure only) by 0.33 \AA.  No reddening correction has been applied.
The error spectrum is overplotted, and principal absorption lines
(mostly interstellar) are identified.  Residual airglow features are
marked with \earth.
\label{b29}
}
\end{figure}

Two successful pointings at Barnard 29 were obtained during the flight,
totaling 5300~s.  The resulting spectra were each rebinned to a common
set of 0.165 \AA\ wavelength bins (about half the instrumental
resolution), then background subtracted, scaled to correct for detector
dead-time effects, and wavelength and flux calibrated as described in
\citeN{ORFEUS98}.  The flux calibration is based on
in-flight observations of the hot DA white dwarf HZ43 and is believed
accurate to about 10\%.  The spectra were weighted by their integration
times and averaged to produce the final calibrated spectrum presented
in \fig{b29}.

\section{ANALYSIS}

\subsection{Atmospheric Parameters}

We begin by deriving an independent set of stellar atmospheric
parameters. To this end, we combine our data with an \iue\ spectrum of
Barnard 29 to produce a complete stellar spectrum extending from $\sim$
2000 \AA\ to the Lyman limit and fit it with the synthetic stellar
spectra of \citeN{Kurucz92}.  The Kurucz stellar atmosphere models
incorporate statistically correct line strengths for some sixty million
atomic and molecular transitions and are available for scaled solar
abundances\footnote{We adopt the usual spectroscopic notations in this
paper, namely $[X] \equiv \log_{10}(X)_{\rm star} -
\log_{10}(X)_{\sun}$ for any quantity $X$, and $\log \epsilon(X) \equiv
\log(N_X/N_{\rm H}) + 12.0$ for absolute number density abundances.}
[M/H] between $+1.0$ and $-5.0$ (where M represents all elements
heavier than helium).  The models provide a thorough treatment of line
blanketing by metals, but their assumption of local thermodynamic
equilibrium (LTE) neglects the non-LTE processes which may be important
for hot, low-gravity stars.

Following \shortciteN{CDK94}, we obtained seven
\iue\ low-resolution spectra of Barnard~29 (SWP 8778, 8779, 9300, 9578,
9600, 11159, and 31824) from the National Space Science Data Center and
averaged them, weighting by the individual exposure times.  The
\iue\ and \orf\ spectra were then binned by 10 \AA\ to match the
resolution of the Kurucz models.  Error bars for each ORFEUS data point
were set to 3\% of the total observed flux (spectrum $+$ background) to
account for fluctuations in the detector response on large spatial
scales \cite{ORFEUS98}.  For consistency, error bars for
the \iue\ data were set in the same way.  The combined spectra were fit
with Kurucz (1992) model spectra using the nonlinear curve-fitting
program SPECFIT \cite{Kriss94} to perform a $\chi^2$ minimization.
Within SPECFIT, models are interpolated to the exact wavelength of each
data point, reddened with a \citeN{CCM89} extinction curve assuming
\ebv\ = 0.02 and \rv\ = 3.1, and scaled by a transmission function to
account for interstellar \hone\ and \ion{O}{1} along the line of sight,
derived from the Bell Labs 21 cm survey \cite{Stark92} according to
the prescription of \citeN{HJD97}.  Free
parameters in the fit are the normalization and wavelength offset of
the model spectrum.

\begin{figure}[tb]
\plotone{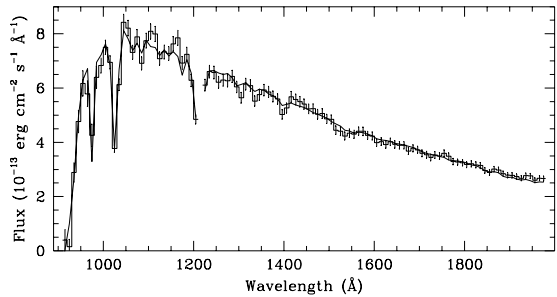}
\caption{Combined \orf\ ($\lambda < 1190$ \AA) and \iue\ spectra of
Barnard 29.  The data have been binned by 10 \AA\ and are shown as a
histogram.  Overplotted is the best-fitting Kurucz (1992) model, with
[M/H] = $-2.0$, $T_{eff}$ = 21,000 K, and $\log g$ = 3.0.  The model has
been reddened with a Cardelli \etal\ (1989) extinction curve assuming
\ebv\ = 0.02 and $R_V$ = 3.1 and includes absorption due to atomic
hydrogen and oxygen assuming a column density of log N(\hone) = 20.26.
\label{kurucz}
}
\end{figure}

The Kurucz model which best fits the data has $T_{eff} = 21,000$ K,
$\log g = 3.0$, and [M/H] = $-2.5$.  The combined spectrum and
best-fitting model are plotted in \fig{kurucz}.  We see that the model
somewhat underpredicts the flux between Lyman $\beta$ and
Lyman $\alpha$.  Whether this discrepancy is due to abundance anomalies
or non-LTE effects must await further analysis.  (For a more complete
discussion of the limitations of this technique, see \citeANP{DDF94}
\citeyearNP{DDF94,DDF95}.)  From the distribution of $\chi^2$ among
the various Kurucz models, we estimate that the uncertainty in the
best-fit parameters is approximately 1000~K in $T_{eff}$, 0.25 dex in
$\log g$, and 0.5 dex in [M/H], though we note that there are no Kurucz
models with $\log g < 3.0$ at the best-fit temperature. These values of
effective temperature and surface gravity are consistent with those
derived by \shortciteN{CDK94} from \iue\ data alone and from
absorption-line fits to the optical spectrum.  The metallicity of the
best-fit Kurucz model is, however, considerably less than that of the
cluster, for which [Fe/H] = $-1.60$ \cite{KSSSLP97}.

To constrain the photospheric abundances of Bar-nard 29, we consider
only the \orf\ spectrum, binned to 0.165 \AA.
We generate high-resolution synthetic stellar spectra using
the stellar atmospheres of Kurucz (1992) and the synthetic spectral
codes of \citeN{Hubeny88}.
In producing these models, we have made extensive use of
the techniques and results of \citeN{BFD96}, which should be
consulted for details of our method.

Adopting the stellar atmospheric parameters and abundances derived by
\shortciteN{CDK94}, we begin with a single stellar model atmosphere
(am15t20000-g30k2.dat; Kurucz 1992) and adjust only the abundances of
elements included in the synthetic spectrum calculation.  We use
SPECFIT to compare a series of models, identical save for the abundance
of the species in question, to the region about a particular line.
Given a grid of such models, SPECFIT interpolates among them to
determine the best-fit abundance, then sets a 1-$\sigma$ error bar on
the result.  The error bar reflects uncertainties both in the continuum
placement and the strength of any nearby lines.

We have determined abundances for all of the elements measured by
\shortciteANP{CDK94} (He, N, O, Al, Si) except Mg, for which our
model predicts no significant absorption in the far UV.  In all cases,
our results are consistent with the optically-determined values, though
our error bars are usually larger, reflecting the lower resolution of
our data.  We will not present those results here.  Instead, we discuss
the three species not available in the optical, C, S, and Fe.

\subsection{Abundances}

\subsubsection{Carbon}

The two strongest carbon lines in the far UV are \ion{C}{3} $\lambda
977$ and \ion{C}{3} $\lambda 1176$.  \shortciteN{BFD96} find that the
\ion{C}{3} $\lambda 1176$ feature is useful for estimating the carbon
abundance, as its strength is relatively insensitive to changes in
temperature and gravity.  The lines are due to an excited-state
transition and thus uncontaminated by interstellar absorption.  Using
this feature, we have determined the star's carbon abundance to be
$6.15 \pm 0.10$ dex [\fig{big_plot}(a)].

\subsubsection{Sulfur}

Our models predict far-UV absorption from three ionization states of
sulfur, \ion{S}{2}, \ion{S}{3}, and \ion{S}{4}.  We have found,
however, that most of the \ion{S}{2} lines yield abundances far lower
than do the \ion{S}{3} and \ion{S}{4} lines.  This would suggest that
the model's effective temperature is too low, but raising $T_{eff}$ by
1000 K, the most allowed by the optical and far-UV fits, does not
relieve the discrepancy, which may instead reflect non-LTE effects not
included in our model atmosphere. We thus use only the high-ionization
lines to constrain the sulfur abundance.  The five strongest \ion{S}{3}
and \ion{S}{4} lines in our bandpass and the abundances or limits
derived from each are
\ion{S}{4} $\lambda 1062.7$, $< 0.95$;
\ion{S}{4} $\lambda \lambda 1073.0, 1073.5$, $5.33 \pm 0.77$;
\ion{S}{3} $\lambda 1077.1$, $6.99 \pm 1.52$;
\ion{S}{3} $\lambda \lambda 1143.6, 1143.9$, $5.41 \pm 0.79$; and
\ion{S}{3} $\lambda 1190.2$, $2.04 \pm 2.02$.
The scatter in these results is due, in part, to line blending with
nearby features (\eg, \ion{Cr}{3} $\lambda 1062.7$, \ion{Fe}{3}
$\lambda 1143.7$, and \ion{Si}{2} $\lambda 1190.4$) and poor fits to
the local stellar continuum.  Ignoring the upper limit, we find a
weighted mean of $\log \epsilon$(S) = $5.34 \pm 0.50$.  Figure
\ref{big_plot}(b) shows the region 1057--1083 \AA\ overplotted by a
synthetic spectrum with our mean sulfur abundance.

\subsubsection{Iron}

A number of strong iron lines are predicted by our models.
Unfortunately, they either lie in regions where the continuum is poorly
fit or are contaminated by strong interstellar absorption.  Instead of
fitting individual features, we have thus chosen to fit a band of iron
lines and the nearby continuum in a well-behaved region of the
spectrum.

\begin{figure}[tbh]
\plotone{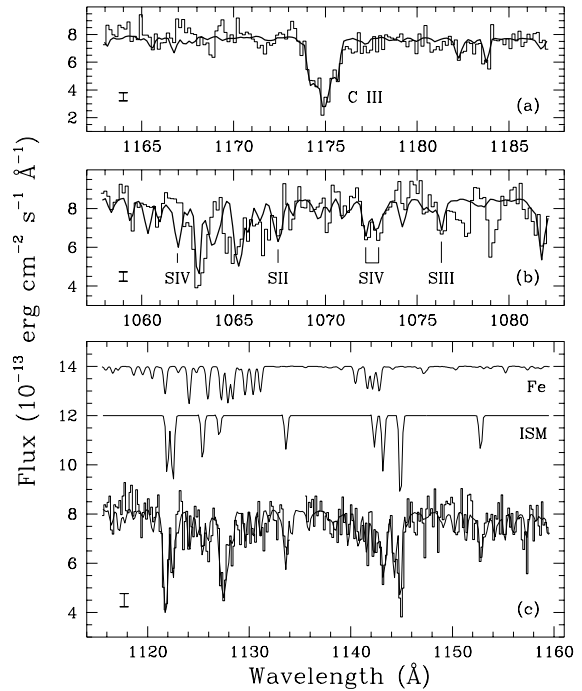}
\caption{(a) The spectrum of Barnard 29 in the region of \ion{C}{3}
$\lambda 1176$ is plotted as a histogram.  The smooth curve represents
our best-fitting model, with a carbon abundance of 6.15.  The mean
1-$\sigma$ uncertainty in this region is indicated at lower left.
(b) Sulfur features in the spectrum of Barnard 29.
A model with the adopted sulfur abundance of 5.34 is overplotted.
(c) Fitting the iron abundance of Barnard 29.
{\it Top curve:}
The ratio of a synthetic stellar spectrum with best-fitting iron
abundance $\log
\epsilon$(Fe) = 5.30 to one with no iron lines, reflecting the absorption
due to iron (mostly Fe III) in the photosphere.
{\it Middle curve:}
ISM features (due to Fe II, Fe III, and P II)
included in the model fit.
The ISM and iron spectra are normalized to the stellar continuum
level, then offset in flux space for clarity.
{\it Bottom curve:}
Histogram is the observed spectrum; smooth line is the best-fitting
model.
\label{big_plot}
}
\end{figure}

Figure \ref{big_plot}(c) shows the spectrum of Barnard 29 between 1115
and 1160 \AA.  (The strong interstellar and airglow features of
\ion{N}{1} $\lambda 1134$ are excluded from the fit.) The iron
spectrum, showing the band between 1120 and 1132 \AA, is plotted at the
top of the figure.  The model includes a synthetic ISM spectrum (also
shown) composed of \ion{Fe}{2}, \ion{Fe}{3}, and \ion{P}{2} lines;
these features are modeled assuming a Doppler parameter b = 10 \kms.
In the fit, the column density of each

\clearpage

\noindent
IS species is allowed to vary
independently.  Synthetic spectra with our best-fit sulfur abundance
exhibit a \ion{S}{2} $\lambda 1124$ feature considerably stronger than
is present in the data, so we have set the sulfur abundance to zero in
these models.  Because the final model contains so many independent
components, we set error bars on the iron abundance by hand, raising
and lowering $\log \epsilon$(Fe) from the best-fit level until $\Delta
\chi^2 = 1$ (corresponding to a 1-$\sigma$ deviation for a single
interesting parameter; \citeNP{Avni76}).  We find $\log
\epsilon$(Fe) = $5.30^{+0.22}_{-0.26}$. The best-fit model is
overplotted.

The best-fit iron abundance is more than 0.5 dex below that of the
Kurucz atmosphere model from which we derive our synthetic spectra.  To
check whether this discrepancy has a significant effect on our result,
we generate a set of three synthetic spectra, with $\log \epsilon$(Fe) =
5.0, 5.5, and 6.0, based on Kurucz model atmospheres with [M/H]
= $-2.5$, $-2.0$, and $-1.5$, respectively.  The models are otherwise the
same as before.  Allowing SPECFIT to interpolate among them,
we find that the best-fit abundance is $\log \epsilon$(Fe) =
$5.28^{+0.26}_{-0.19}$, consistent with our previous result.  We
conclude that a small discrepancy in the metallicity of the input
stellar atmosphere model is not a significant source of error in our
derived abundances.

\section{DISCUSSION}

The CNO abundance pattern seen in Barnard 29 is observed in a number of
red giants in M13, four carbon-poor planetary nebulae, and all of the
high-latitude B-type post-AGB candidates \shortcite{CDK94}.  According
to current stellar evolutionary theory, such abundance ratios are
expected of stars that leave the AGB before third dredge up brings
significant nuclear-processed material to the surface \cite{IR83}.  We
would therefore expect Barnard 29 to have an iron abundance similar to
other M13 giants, for which [Fe/H] = $-1.60$ \cite{KSSSLP97}.
Instead, we find that the star has an iron abundance 2 dex or more
below solar.  Low metallicities are seen in other UV-bright stars in
globular clusters, but not all: for vZ 1128 in M3, [M/H] = $-3.5 \pm
1.5$, nearly 2 dex below the cluster mean, while BS in 47 Tuc and UV5
in NGC 1851 have metallicities consistent with the cluster mean
(\shortciteANP{DDF94} \citeyearNP{DDF94,DDF95}).

For BD$+33$\arcdeg 2642, a planetary nebula central star in the
galactic halo, \citeN{NHK94} derive an abundance distribution similar
to that of Barnard 29: He is near solar, C, N, O, Mg, Si are depleted
by about 1 dex, and Fe is depleted by 2 dex relative to the sun.  The
authors suggest that the low Fe abundance may be the result of gas-dust
fractionation, a process proposed to explain the chemical peculiarities
in some cooler post-AGB stars.  In this scenario, metals with a high
condensation temperature condense into dust grains and are removed by
radiation pressure, while elements with lower condensation temperatures
remain.  Most models require a binary system to support a circumstellar
disk about the AGB star (for a review, see \citeNP{TWW93p103}), but a
single-star model has been proposed by \citeN{ML92}.  One
of its predictions, however, is that $N$(C) $\approx$ $N$(O) in the
resulting iron-poor photosphere, a condition not seen in Barnard 29.

The abundances presented here are derived under the assumption of LTE.
Because Barnard 29 is a low-gravity star, non-LTE effects may be
significant.  The strength of such effects and their influence on the
derived stellar abundances are difficult to determine {\it a priori}.
Future work will address these issues using appropriate non-LTE model
atmospheres.

\acknowledgments

This research has made use of the NASA ADS Abstract Service and the
Catalogue Service of the CDS, Strasbourg, France. We thank R.\ Kurucz
for providing a computer-readable tape of his stellar atmosphere
models.  We thank I.\ Hubeny for providing his spectral synthesis codes
and T.\ Brown for assistance in using them.  We acknowledge our
colleagues on the \orf\ team and the many NASA and DARA personnel who
helped make the \orf\ mission successful. This work is supported by
NASA grant NAG5-696.


\clearpage 





\end{document}